\documentclass[onecolumn, draftcls, 12pt]{IEEEtran}

\usepackage{amsfonts}
\usepackage{mathrsfs}
\usepackage{amssymb,amsmath}

\usepackage{algorithm}
\usepackage{algorithmic}
\usepackage{amsmath,amssymb,epsfig,graphics,subfigure}
\usepackage{theorem}
\usepackage{array,color}
\usepackage[compress]{cite}

\theoremheaderfont{\normalfont\bfseries}

\begin{document}

\title{A Matrix-Field Weighted Mean-Square-Error Model for MIMO Transceiver Designs}

\author{Chengwen Xing, Wenzhi Li, Shaodan Ma, Zesong Fei and Jingming Kuang
\thanks{C. Xing, W. Li, Z. Fei and J. Kuang are with the School of Information and Electronics, Beijing Institute of Technology, Beijing, 100081, China. (E-mail: chengwenxing@ieee.org, wenzhi306@163.com, \{feizesong,jmkuang\}@bit,edu.cn).
S. Ma is with the Department of Electrical and Computer Engineering, University of Macau, Macao (email:
shaodanma@umac.mo).
This work was supported in part by National Natural Science Foundation of China under Grant No. 61101130.} }
\maketitle

\begin{abstract}
In this letter, we investigate an important and  famous issue, namely weighted mean-square-error (MSE) minimization transceiver designs. In our work, for transceiver designs a novel weighted MSE model is proposed, which is defined as a linear matrix function with respect to the traditional data detection MSE matrix. The new model can  be interpreted an extension of weighting operation from vector field to matrix field. Based on the proposed weighting operation a general transceiver design is proposed, which aims at minimizing an increasing matrix-monotone function of the output of the previous linear matrix function. The structure of the optimal solutions is also derived. Furthermore, two important special cases of the matrix-monotone functions are discussed in detail. It is also revealed  that these two problems are exactly equivalent to the transceiver designs of sum MSE minimization and capacity maximization for dual-hop amplify-and-forward  (AF) MIMO relaying systems, respectively. Finally, it is concluded that the AF relaying is undoubtedly this kind of weighting operation.
\end{abstract}

\begin{keywords}
Transceiver deigns, weighted MSE, MIMO.
\end{keywords}

\section{Introduction}
\label{sect:intro}

It is well-established that multiple-input multiple-output (MIMO) technology  is a powerful transmission technology due to its capability to bring both spatial diversity and multiplexing gains to wireless communication systems \cite{Larsson03}.
In order to realize the promised performance gains of MIMO systems, MIMO transceiver designs are usually of great importance. Referring to MIMO transceiver designs, there are various objective functions such as capacity, mean-square-error (MSE), signal-to-interference-plus-noise-ratio (SINR), bit error rate (BER) and so on \cite{Sampth01,Palomar03}. The various performance metrics reflect different design preferences \cite{Sampth01,Palomar03}.

The simplest way to unify  various performance metrics into a single objective function is to adopt weighted MSE minimization, i.e, minimizing weighted sum of the diagonal elements of the data detection MSE at the destination \cite{Sampth01}. Unfortunately, in most cases the simplest one may not be the best one. After the landmark paper \cite{Sampth01}, it was revealed in \cite{Palomar03} that for certain performance metrics e.g., BER (Schur-convex function of the diagonal elements of the MSE matrix), the optimal solution may have a different structure  from  that of the weighted MSE minimization discussed in \cite{Sampth01}. A natural question is whether we can improve the effectiveness of the weighting operation. This is the motivation of our work.

In this paper, we investigate the weighted MSE minimization transceiver designs and try to improve the existing weighted MSE model. Different from the fact that the traditional MSE weighting operation for transceiver designs only considers  weighted sum of the diagonal elements of the data detection MSE matrix \cite{Sampth01,Palomar03}, in our work a novel weighted MSE model is proposed in which the weighting operation is defined as a general linear matrix function with respect to the data detection MSE matrix. The output of the linear matrix function is named as weighted MSE matrix. Based on the new model, a general transceiver design of minimizing an increasing matrix-monotone function of the weighted MSE matrix is proposed. Generally speaking, this procedure is an extension of weighting operation from vector field to matrix field.

The structure of the optimal solutions for the general matrix-monotone objective functions is derived first and then two important special cases are investigated in more detail. Furthermore, it is revealed that these two cases have clear physical meanings as they are exactly equivalent to the transceiver designs of sum MSE minimization and capacity maximization for dual-hop amplify-and-forward (AF) MIMO relaying systems, respectively. Finally, an interesting conclusion is drawn that the AF relaying is exactly this weighting operation. The authors think this may be the reason why linear transceiver designs for AF MIMO relaying systems can enjoy almost all the results for point-to-point systems.

\noindent \textbf{Notation:}  The notation ${\bf{Z}}^{\rm{H}}$ denotes the
Hermitian transpose of the matrix ${\bf{Z}}$, and ${\rm{Tr}}({\bf{Z}})$ is the
trace of the matrix ${\bf{Z}}$. The notation ${\bf{Z}}^{1/2}$ is the
Hermitian square root of the positive semi-definite matrix
${\bf{Z}}$. For a rectangular diagonal matrix ${\boldsymbol \Lambda}$, ${\boldsymbol \Lambda} \searrow$ denotes the main diagonal elements are in decreasing order and ${\boldsymbol \Lambda} \nearrow$ denotes the main diagonal elements are in increasing order. For two Hermitian matrices, ${\bf{C}} \succeq
{\bf{D}}$ means that ${\bf{C}}-{\bf{D}}$ is a positive semi-definite
matrix. The symbol $\lambda_i({\bf{Z}})$ represents the $i^{\rm{th}}$ largest eigenvalue of ${\bf{Z}}$.

\section{Motivations and Problem Formulation}
 We embark on our work by investigating a point-to-point MIMO system in which there is one source and one destination. Both nodes are equipped with multiple antennas.  Without loss of generality, the numbers of transmit antennas and receive antennas are $N_{\rm{Tx}}$ and $N_{\rm{Rx}}$, respectively. In signal transmission period, the signal model is given as
\begin{align}
{\bf{y}}={\bf{H}}{\bf{F}}{\bf{s}}+{\bf{n}}
\end{align}where ${\bf{y}}$ is the $N_{\rm{Rx}}\times 1$ received signal at the destination. The matrix ${\bf{H}}$ is the $N_{\rm{Rx}}\times N_{\rm{Tx}}$ channel matrix between the transmitter and the receiver. Moreover, ${\bf{s}}$ is the $N_{\rm{Dat}}\times 1$ transmitted signal with identity covariance matrix, i.e., ${\mathbb E}\{{\bf{s}}{\bf{s}}^{\rm{H}}\}={\bf{I}}$ and ${\bf{F}}$ is the precoding matrix at the transmitter. Finally, the $N_{\rm{Rx}}\times 1$ vector ${\bf{n}}$ denotes the additive noise vector at the receiver with mean zero and covariance matrix of ${\bf{R}}_n$. This is an extensively studied linear model.

With linear equalizer ${\bf{G}}$, the mean square error (MSE) matrix of data detection at the destination equals
\begin{align}
\label{MSE_Matrix}
{\boldsymbol \Phi}({\bf{G}},{\bf{F}})= \mathbb{E}\{({\bf{G}}{\bf{y}}-{\bf{s}})
({\bf{G}}{\bf{y}}-{\bf{s}})^{\rm{H}}\},
\end{align}which is called \textbf{original MSE matrix} in the remaining part of this paper. For the traditional designs, the optimization problem of minimizing weighted MSE is formulated as \cite{Sampth01}
\begin{align}
\label{class_opt}
& \min \ \ \ \ {\sum}_j w_j[{\boldsymbol \Phi}({\bf{G}},{\bf{F}})]_{j,j} \nonumber \\ & \  {\rm{s.t.}} \ \ \ \ \  {\rm{Tr}}({\bf{F}}{\bf{F}}^{\rm{H}})\le P,
\end{align}where $[{\bf{Z}}]_{j,j}$ denotes the $(i,j)^{\rm{th}}$ element of matrix ${\bf{Z}}$ and $w_j$'s are the weighting factors. In addition, $P$ is the maximum transmit power at the transmitter. The optimization problem (\ref{class_opt}) covers sum MSE minimization and capacity maximization as its special cases \cite{Sampth01}. However, Palomar pointed out that for certain performance metric such as BER minimization, the optimal solution is different from that of problem (\ref{class_opt}) \cite{Palomar03}. A question is whether we can develop a more powerful weighting operation to cover more cases.

It is obvious that in the traditional weighting operation the off-diagonal entries of the original MSE matrix are always neglected, and some information may be lost. To overcome this problem, we try to develop a new weighting operation which is a function of all entries of the original MSE matrix and can also effectively reshape the original MSE matrix ${\boldsymbol \Phi}({\bf{G}},{\bf{F}})$. As we are focusing on a linear operation on ${\boldsymbol \Phi}({\bf{G}},{\bf{F}})$, the most general operation is obviously a linear matrix function with respect to ${\boldsymbol \Phi}({\bf{G}},{\bf{F}})$ and its paramount feature is that the output is also a matrix \cite{Marshall79}. In other words, the proposed weighting operation aims at generating a new matrix to reflect the designer's preference. In the following, the output matrix is named as \textbf{weighted MSE matrix} to distinguish from the former \textbf{original MSE matrix}. This can be understood as an extension of weighting operation from vector field to matrix field. The one in matrix field usually owns more information than its counterpart in vector field.

Before designing the weighting operation, at the beginning we list several desired properties to be met.

\noindent $\bullet$ \textbf{\textsl{ The output matrix must also be a positive semi-definite Hermitian matrix, as this is the most important property of an MSE matrix or a covariance matrix.}}

\noindent $\bullet$ \textbf{\textsl{The new weighting operation must be able to cover the classical one as it special case. If it is more powerful, it definitely covers the weaker one as its special case.}}

\noindent $\bullet$ \textbf{\textsl{The operation should better be a linear operation on the MSE matrix in (\ref{MSE_Matrix}). It is because weighting is definitely a linear operation.}}

 In order to meet the three requirements, a natural  choice of linear weighting operation over the original MSE matrix is defined as
\begin{align}
\label{New_Weight}
{\boldsymbol \Psi}({\bf{G}},{\bf{F}})={\sum}_{k=1}^K{\bf{W}}_k^{\rm{H}}{\boldsymbol \Phi}({\bf{G}},{\bf{F}}){\bf{W}}_k+{\boldsymbol{\Pi}}
\end{align}where ${\bf{W}}_k$ is a complex matrix (which is not limited to square matrices) and ${\boldsymbol{\Pi}}$ is a positive semi-definite matrix. Both ${\bf{W}}_k$'s and ${\boldsymbol{\Pi}}$ are constant. It is obvious that (\ref{New_Weight}) owns all of the previously listed properties. In the following, ${\boldsymbol \Psi}({\bf{G}},{\bf{F}})$ is referred to  as the \textbf{weighted MSE matrix}.

Based on the proposed weighting operation, the optimization problem of transceiver designs is formulated as follows
\begin{align}
\label{Opt_0}
&\min_{{\bf{G}},{\bf{F}}} \ \ \ {f}[{\boldsymbol \Psi}({\bf{G}},{\bf{F}})] \nonumber¡¡\\
& \ {\rm{s.t.}} \ \ \ \ {\rm{Tr}}({\bf{F}}{\bf{F}}^{\rm{H}})\le P,
\end{align}where $f(\bullet)$ is an increasing matrix-monotone function, i.e., ${\boldsymbol A} \preceq {\boldsymbol B}$ $\rightarrow$ $f({\boldsymbol A}) \le f({\boldsymbol B})$ \cite{Marshall79}. In the following section, the structure of optimal solution of (\ref{Opt_1}) will be investigated first.

\section{The Structure of Optimal Solutions}

It is well-known that linear minimum mean-square-error (LMMSE) equalizer is the dominating equalizer for linear equalizers \cite{Kay93}, which is equivalent to
\begin{align}
\label{LMMSE_Euq}
{\bf{G}}_{\rm{LM}}=({\bf{H}}{\bf{F}})^{\rm{H}}({\bf{H}}{\bf{F}}{\bf{F}}
^{\rm{H}}{\bf{H}}^{\rm{H}}+{\bf{R}}_{n})^{-1}.
\end{align}It means that for any linear equalizer ${\bf{G}}$, the LMMSE equalizer has the  property
${\boldsymbol \Phi}({\bf{G}}_{\rm{LM}},{\bf{F}})\preceq {\boldsymbol \Phi}({\bf{G}},{\bf{F}})$ based on which we have the following relationship
\begin{align}
\underbrace{\sum_{k=1}^K{\bf{W}}_k^{\rm{H}}{\boldsymbol \Phi}({\bf{G}}_{\rm{LM}},{\bf{F}}){\bf{W}}_k+{\boldsymbol \Pi}}_{{\boldsymbol \Psi}({\bf{G}}_{\rm{LM}},{\bf{F}})}\preceq \underbrace{\sum_{k=1}^K{\bf{W}}_k^{\rm{H}}{\boldsymbol \Phi}({\bf{G}},{\bf{F}}){\bf{W}}_k+{\boldsymbol \Pi}}_{{\boldsymbol \Psi}({\bf{G}},{\bf{F}})}.
\end{align}As a result  it can be concluded that LMMSE equalizer is exactly the optimal solution of ${\bf{G}}$. Substituting (\ref{LMMSE_Euq}) into the original MSE matrix (\ref{MSE_Matrix}), we directly have
\begin{align}
&{\boldsymbol \Phi}({\bf{G}}_{\rm{LM}},{\bf{F}}) =(
{\bf{F}}^{\rm{H}}{\bf{H}}^{\rm{H}}{\bf{R}}_n^{-1}{\bf{H}}{\bf{F}}+
{\bf{I}})^{-1},
\end{align} based on which the weighted MSE matrix (\ref{New_Weight}) is equal to
\begin{align}
\label{Weighted_MSE}
{\boldsymbol \Psi}({\bf{G}}_{\rm{LM}},{\bf{F}})
=&{\sum}_{k=1}^K{\bf{W}}_k^{\rm{H}}(
{\bf{F}}^{\rm{H}}{\bf{H}}^{\rm{H}}{\bf{R}}_n^{-1}{\bf{H}}{\bf{F}}+
{\bf{I}})^{-1}{\bf{W}}_k+{\boldsymbol \Pi} \nonumber \\
\triangleq & {\boldsymbol \Psi}({\bf{F}}).
\end{align}
Therefore, based on (\ref{Weighted_MSE}) the optimization problem (\ref{Opt_0}) becomes
\begin{align}
\label{Opt_Original}
&\min_{{\bf{F}}} \ \ \ {f}[{\boldsymbol \Psi}({\bf{F}})]=g({\bf{F}}^{\rm{H}}{\bf{H}}^{\rm{H}}{\bf{R}}_{n}^{-1}{\bf{H}}{\bf{F}}) \nonumber¡¡\\
& \ {\rm{s.t.}} \ \ \ {\rm{Tr}}({\bf{F}}{\bf{F}}^{\rm{H}})\le P.
\end{align}
 Notice that for the objective function $g(\bullet)$ is a decreasing matrix-monotone function \cite{XingTSP2013}, and thus  it can be proved that the structure of the optimal ${\bf{F}}$ has the following structure \cite{XingTSP2013}.

\noindent \underline{\textbf{Conclusion 1:}} Defining the following singular value decomposition, ${\bf{R}}_n^{-1/2}{\bf{H}}=
{\bf{U}}_{\boldsymbol{\mathcal{H}}}
{\boldsymbol\Lambda}_{\boldsymbol{\mathcal{H}}}{\bf{V}}_{\boldsymbol{\mathcal{H}}}^{\rm{H}}
$ with ${\boldsymbol\Lambda}_{\boldsymbol{\mathcal{H}}} \searrow$,
the optimal solution of ${\bf{F}}$ of problem (\ref{Opt_Original}) has the following structure
\begin{align}
\label{F_OPT}
{\bf{F}}_{\rm{opt}}={\bf{V}}_{\boldsymbol{\mathcal{H}}}{\boldsymbol \Lambda}_{\bf{F}}{\bf{U}}_{\bf{F}}^{\rm{H}} \ \ \text{with} \ \ {\boldsymbol \Lambda}_{\bf{F}}^{\rm{H}}{\boldsymbol\Lambda}_{\boldsymbol{\mathcal{H}}}^{\rm{H}}{\boldsymbol\Lambda}_{\boldsymbol{\mathcal{H}}}{\boldsymbol \Lambda}_{\bf{F}} \searrow
\end{align}where ${\boldsymbol \Lambda}_{\bf{F}}$ is a rectangular diagonal matrix and ${\bf{U}}_{\bf{F}}$ is a unitary matrix.

Note that the specific formulations of ${\boldsymbol \Lambda}_{\bf{F}}$ and ${\bf{U}}_{\bf{F}}$ are determined by the the specific functions of $f(\bullet)$, and must be investigated case by case. We can only give a common logic for the derivations instead of detailed solutions and in the following, two special cases are investigated to show this logic.

\section{Two Special Cases}

In this section, we focus on a simple but representative special case of $K=1$ with ${\bf{W}}_1={\bf{W}}$. As for any Hermitian matrix in general there are two important parameters i.e.,  trace and determinant, in this section two special cases of  optimization problem (\ref{Opt_Original}) are investigated, which aim at minimizing the trace and the determinant of the weighted MSE matrix ${\boldsymbol \Psi}({\bf{F}})$, respectively. Notice that the two objective functions are obviously increasing matrix-monotone functions and then their optimal solutions satisfy \textbf{Conclusion 1}. The remaining problem is how to derive ${\bf{U}}_{\bf{F}}$
and  ${\boldsymbol \Lambda}_{\bf{F}}$
and this is exactly the main focus of this section.

\subsection{Minimization the trace of ${\boldsymbol \Psi}({\bf{F}})$}
For minimizing the trace of the weighted MSE matrix, the optimization problem (\ref{Opt_Original}) becomes
\begin{align}
\label{P_1}
{\textbf{P 1:}} \ \  &  {\min_{\bf{F}}} \ \ \ {f}[{\boldsymbol \Psi}({\bf{F}})]={\rm{Tr}}[{\boldsymbol \Psi}({\bf{F}})] \nonumber \\
& \ {\rm{s.t.}} \ \ \ {\rm{Tr}}({\bf{F}}{\bf{F}}^{\rm{H}}) \le P.
\end{align}
Based on the formulation of ${\boldsymbol \Psi}({\bf{F}})$ given in (\ref{Weighted_MSE}), the previous optimization problem is equivalent to the following one
\begin{align}
\label{Opt_1}
& {\min_{\bf{F}}} \ \ \ {\rm{Tr}}[{\bf{W}}^{\rm{H}}(
{\bf{F}}^{\rm{H}}{\bf{H}}^{\rm{H}}{\bf{R}}_n^{-1}{\bf{H}}{\bf{F}}+{\bf{I}})^{-1}{\bf{W}} ]+{\rm{Tr}}({\boldsymbol \Pi}) \nonumber \\
& \ {\rm{s.t.}} \ \ \ {\rm{Tr}}({\bf{F}}{\bf{F}}^{\rm{H}}) \le P.
\end{align}
Before discussing the minimum value of the above function, a famous matrix inequality is first given.

\noindent \textbf{Inequality 1:} The trace of a product of two $N \times N$ positive semi-definite matrices has the following inequality \cite{Marshall79}
\begin{align}
& {\textbf{Inequ 1:}} \ \ {\sum}_{i=1}^N\lambda_i({\boldsymbol A})\lambda_{N-i+1}({\boldsymbol B})  \le {\rm{Tr}}({\boldsymbol A}{\boldsymbol B}) \label{inequ_1},
\end{align}where the equality holds when
${\bf{U}}_{{\boldsymbol {A}}}={\bf{\bar U}}_{{\boldsymbol {B}}}^{\rm{H}}$ with ${\bf{U}}_{{\boldsymbol {A}}}$ and ${\bf{\bar U}}_{{\boldsymbol {B}}}^{\rm{H}}$ defined based on the eigenvalue decompostions
${\boldsymbol A}={\bf{U}}_{{\boldsymbol {A}}}{\boldsymbol \Lambda}_{{\boldsymbol {A}}}{\bf{U}}_{{\boldsymbol {A}}}^{\rm{H}} $ with $ {\boldsymbol \Lambda}_{{\boldsymbol {A}}} \searrow$ and
${\boldsymbol B}={\bf{\bar U}}_{{\boldsymbol { B}}}{\boldsymbol {\bar \Lambda}}_{{\boldsymbol {B}}}{\bf{\bar U}}_{{\boldsymbol {B}}}^{\rm{H}}$  with  ${\boldsymbol{\bar  \Lambda}}_{{\boldsymbol {B}}} \nearrow$.

Based on (\ref{F_OPT}) and \textbf{Inequ 1}, the optimal solution of ${\bf{U}}_{\bf{F}}$ in (\ref{F_OPT}) for \textbf{P 1} has the following property.

\noindent \underline{\textbf{Conclusion 2:}} For the trace minimization problem (\ref{P_1}), the unitary matrix ${\bf{U}}_{\bf{F}}$ of the optimal solution given by (\ref{F_OPT}) satisfies
\begin{align}
{\bf{U}}_{\bf{F}}={\bf{U}}_{\bf{W}}
\end{align}where ${\bf{U}}_{\bf{W}}$ is the unitary matrix defined based on the following singular value decomposition ${\bf{W}}={\bf{U}}_{\bf{W}}{\boldsymbol \Lambda}_{\bf{W}}{\bf{V}}_{\bf{W}}^{\rm{H}}$  with ${\boldsymbol \Lambda}_{\bf{W}} \searrow$.

Based on \textbf{Conclusions 1} and \textbf{2}, the remaining variable is only ${\boldsymbol \Lambda}_{\bf{F}}$. Denoting $[{\boldsymbol \Lambda}_{\bf{W}}]_{j,j}=\lambda_{w,j}^{1/2}$, $[{\boldsymbol \Lambda}_{\boldsymbol{\mathcal{H}}}]_{j,j}=\lambda_{h,j}^{1/2}$ and  $[{\boldsymbol \Lambda}_{\bf{F}}]_{j,j}=f_{j}$ the optimization problem (\ref{Opt_1}) is simplified as
\begin{align}
& \min_{f_j^2} \ \ \  {\sum}_j \frac{\lambda_{w,j}}{1+\lambda_{h,j}f_{j}^2}, \ \ \ \ {\rm{s.t.}} \ \ {{\sum}_j} f_{j}^2 \le P
\end{align}whose optimal solution is obviously water-filling \cite{Boyd04}.

\noindent \textbf{Remark:} If ${\bf{U}}_{\bf{W}}={\bf{U}}_{\rm{DFT}}$ where ${\bf{U}}_{\rm{DFT}}$ is the discrete Fourier transform (DFT) matrix and the diagonal elements of ${\boldsymbol \Lambda}_{\bf{W}}$ are only slightly different (almost the same), the optimal solution is exactly the optimal  solution for Schur-convex objective functions given in\cite{Palomar03}.

\subsection{Minimize the determinant of ${\boldsymbol \Psi}({\bf{F}})$}

On the other hand, the optimization problem minimizing the determinant of ${\boldsymbol \Psi}({\bf{F}})$ reads as
\begin{align}
\label{P_2}
{\textbf{P 2:}} \ \ & \min_{\bf{F}} \ \ {f}[{\boldsymbol \Psi}({\bf{F}})]={\rm{log}}|{\boldsymbol \Psi}({\bf{F}})| \nonumber \\
& \ {\rm{s.t.}} \ \ \ {\rm{Tr}}({\bf{F}}{\bf{F}}^{\rm{H}}) \le P.
\end{align}Based on the definition of ${\boldsymbol \Psi}({\bf{F}})$, \textbf{P 2} is obviously equivalent to
\begin{align}
\label{Opt_2}
& \min_{\bf{F}} \ \ {\rm{log}}|{\boldsymbol \Pi}+{\bf{W}}^{\rm{H}}(
{\bf{F}}^{\rm{H}}{\bf{H}}^{\rm{H}}{\bf{R}}_n^{-1}{\bf{H}}{\bf{F}}+{\bf{I}})^{-1}{\bf{W}} | \nonumber \\
& \ {\rm{s.t.}} \ \ \ {\rm{Tr}}({\bf{F}}{\bf{F}}^{\rm{H}}) \le P.
\end{align}
The objective function of (\ref{Opt_2}) can be further reformulated as
\begin{align}
\label{Object_P_2}
&{\rm{log}}|{\bf{W}}^{\rm{H}}(
{\bf{F}}^{\rm{H}}{\bf{H}}^{\rm{H}}{\bf{R}}_{n}^{-1}
{\bf{H}}{\bf{F}}+{\bf{I}})^{-1}{\bf{W}}+{\boldsymbol{\Pi}}|\nonumber \\
=&{\rm{log}}(|{\boldsymbol{\Pi}}||{\bf{W}}^{\rm{H}}({\bf{F}}^{\rm{H}}
{\bf{H}}^{\rm{H}}{\bf{R}}_{n}^{-1}
{\bf{H}}{\bf{F}}+{\bf{I}})^{-1}{\bf{W}}{{\boldsymbol{\Pi}}}^{-1}
+{\bf{I}}|) \nonumber \\
=&{\rm{log}}|{\boldsymbol{\Pi}}|+{\rm{log}}|({\bf{F}}^{\rm{H}}
{\bf{H}}^{\rm{H}}{\bf{R}}_{n}^{-1}
{\bf{H}}{\bf{F}}+{\bf{I}})^{-1}{\bf{W}}{\boldsymbol{\Pi}}^{-1}{\bf{W}}^{\rm{H}}
+{\bf{I}}| \nonumber \\
=&{\rm{log}}|{\boldsymbol{\Pi}}|+{\rm{log}}|
{\bf{W}}{\boldsymbol{\Pi}}^{-1}{\bf{W}}^{\rm{H}}+({\bf{F}}^{\rm{H}}{\bf{H}}^{\rm{H}}
{\bf{R}}_{n}^{-1}
{\bf{H}}{\bf{F}}+{\bf{I}})|\nonumber \\
&-{\rm{log}}|({\bf{F}}^{\rm{H}}{\bf{H}}^{\rm{H}}{\bf{R}}_{n}^{-1}
{\bf{H}}{\bf{F}}+{\bf{I}})|,
\end{align}where the second equality is based on the fact that $|{\boldsymbol {A}}{\boldsymbol {B}}+{\bf{I}}|=|{\boldsymbol {B}}{\boldsymbol {A}}+{\bf{I}}|$.
The further derivation of the minimum value of (\ref{Object_P_2}) relies on the following inequality.

\noindent \textbf{Inequality 2:} The determinant of the sum of two positive semi-definite matrices has the following inequality
\begin{align}
& {\textbf{Inequ 2:}} \ \ {\prod}_{i=1}^N(\lambda_i({\boldsymbol A})+\lambda_{i}({\boldsymbol B}))  \le |{\boldsymbol A}+{\boldsymbol B}|, \label{inequ_2} \end{align}where the equality holds when
${\bf{U}}_{{\boldsymbol {A}}}={\bf{U}}_{{\boldsymbol {B}}}$ and the unitary matrix ${\bf{U}}_{{\boldsymbol {B}}}$ is defined based on
${\boldsymbol B}={\bf{ U}}_{{\boldsymbol { B}}}{\boldsymbol \Lambda}_{{\boldsymbol {B}}}{\bf{ U}}_{{\boldsymbol {B}}}^{\rm{H}}$ with ${\boldsymbol{  \Lambda}}_{{\boldsymbol {B}}} \searrow$.

Based on {\textbf{Inequ 2}}, the objective function (\ref{Object_P_2}) satisfies
\begin{align}
\label{case_b}
& {\rm{log}}|{\bf{W}}^{\rm{H}}(
{\bf{F}}^{\rm{H}}{\bf{H}}^{\rm{H}}{\bf{R}}_{n}^{-1}
{\bf{H}}{\bf{F}}+{\bf{I}})^{-1}{\bf{W}}+{\boldsymbol{\Pi}}|
  \ge\nonumber \\&{\rm{log}}|{\boldsymbol{\Pi}}|+ \sum_i{\rm{log}}[\lambda_i({\bf{W}}{\boldsymbol{\Pi}}^{-1}{\bf{W}}^{\rm{H}})
+\lambda_i({\bf{F}}^{\rm{H}}{\bf{H}}^{\rm{H}}{\bf{R}}_{n}^{-1}
{\bf{H}}{\bf{F}}+{\bf{I}})]\nonumber \\
&-\sum_i{\rm{log}}[\lambda_i({\bf{F}}^{\rm{H}}{\bf{H}}^{\rm{H}}{\bf{R}}_{n}^{-1}
{\bf{H}}{\bf{F}}+{\bf{I}})].
\end{align} In order to make the equality in (\ref{case_b}) hold, based on \textbf{Conclusion 1} and the inequality (\ref{inequ_2}) we have the following conclusion.

\noindent \underline{\textbf{Conclusion 3:}} Defining the eigenvalue  decomposition, $
{\bf{W}}{\boldsymbol{\Pi}}^{-1}{\bf{W}}^{\rm{H}}=
{\bf{U}}_{{\boldsymbol \Theta}}
{\boldsymbol\Lambda}_{{\boldsymbol \Theta}}{\bf{U}}_{{\boldsymbol \Theta}}^{\rm{H}}$ with ${\boldsymbol\Lambda}_{{\boldsymbol \Theta}} \searrow
 $, for the determinant  minimization problem (\ref{P_2}), the unitary matrix ${\bf{U}}_{\bf{F}}$ of the optimal solution given by (\ref{F_OPT}) is equal to
\begin{align}
{\bf{U}}_{\bf{F}}={\bf{U}}_{{\boldsymbol \Theta}}.
\end{align}

Based on \textbf{Conclusions 1} and \textbf{3}, and defining $[{\boldsymbol\Lambda}_{{\boldsymbol \Theta}}]_{j,j}=\lambda_{\Theta,j}$, the optimization problem (\ref{P_2}) can be  rewritten as
\begin{align}
& \min_{f_j^2} \ \ \  {\sum}_j {\rm{log}}\left(\frac{\lambda_{\Theta,j}}{\lambda_{h,j}f_{j}^2+1}+1\right), \ \ {\rm{s.t.}} \ \ {{\sum}_j} f_{j}^2 \le P
\end{align}whose the optimal solution is also water-filling \cite{XingTSP2013}.

\section{The Physical Meanings}

\begin{figure}[!ht]
\centering
\includegraphics[width=.8\textwidth]{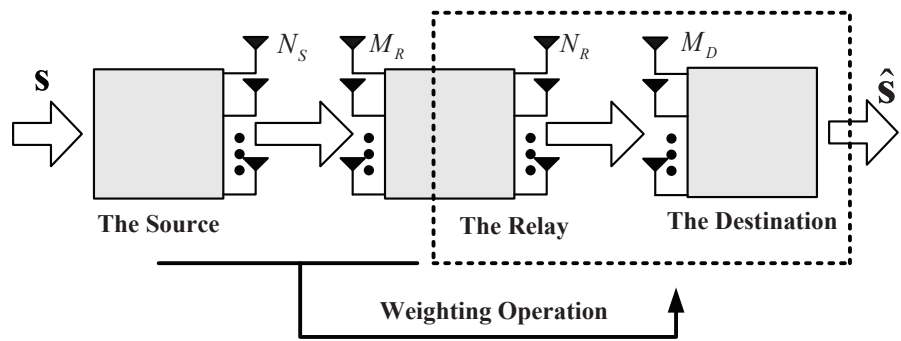}
\caption{The amplify-and-forward MIMO relaying diagram.}\label{fig:1}
\end{figure}

In this section, a dual-hop amplify-and-forward (AF) cooperative
communication system is presented to show the physical meanings of the previously discussed optimization problems \textbf{P 1} and \textbf{P 2}. In the considered AF relaying system, there is one source, one relay, and one destination which are all equipped with multiple antennas as shown in Fig.~\ref{fig:1}.
At the first hop, the source
transmits signal to the relay and the signal vector is denoted as ${\bf{s}}$ with the covariance matrix
${\bf{R}}_{s}=\mathbb{E}\{{\bf{s}}{\bf{s}}^{\rm{H}}\}$. The matrix
${\bf{H}}_{1}$ represents the MIMO channel matrix between the source and
relay. The symbol ${\bf{n}}_1$ represents the additive Gaussian noise at the relay with covariance matrix ${\bf{R}}_{n_1}$. At the relay, the received signal is multiplied by a forwarding matrix ${\bf{P}}$ and then directly forwarded to the destination without decoding. Similarly, the MIMO channel matrix between the relay and destination is denoted as ${\bf{H}}_{2}$, and the additive Gaussian noise vector at the destination is denoted as ${\bf{n}}_2$ with covariance matrix ${\bf{R}}_{n_2}$.

Based on the previous definitions, the parameters involved in \textbf{P 1} and \textbf{P 2} are firstly set to be
\begin{align}
&{\bf{H}}={\bf{H}}_2,  \ \ {\bf{R}}_{n}= {\bf{R}}_{n_2}, \ \ K=1, \nonumber \\
& {\bf{W}}_1=({\bf{H}}_1{\bf{R}}_s{\bf{H}}_1^{\rm{H}}+{\bf{R}}_{n_1})^{-1/2}
{\bf{H}}_1{\bf{R}}_s, \nonumber \\
& {\boldsymbol \Pi}={\bf{R}}_s-{\bf{W}}_1^{\rm{H}}{\bf{W}}_1, \ \ {\bf{F}}= {\bf{P}}({\bf{H}}_1{\bf{R}}_s{\bf{H}}_1^{\rm{H}}+{\bf{R}}_{n_1})^{1/2}.
\end{align}
Then the \textbf{weighted MSE matrix} ${\boldsymbol \Psi}({\bf{F}})$ in (\ref{Weighted_MSE}) becomes to be
\begin{align}
\label{New_MSE_Matrix}
{\boldsymbol \Psi}({\bf{F}})&={\bf{R}}_s-{\bf{R}}_s{\bf{H}}_1^{\rm{H}}{\bf{P}}^{\rm{H}}{\bf{H}}_2^{\rm{H}}[{\bf{H}}_2{\bf{P}}
({\bf{H}}_1{\bf{R}}_s{\bf{H}}_1^{\rm{H}}+{\bf{R}}_{n_1})\nonumber \\
& \times {\bf{P}}^{\rm{H}}{\bf{H}}_2^{\rm{H}}+{\bf{R}}_{n_2}]^{-1}{\bf{H}}_2{\bf{P}}{\bf{H}}_1{\bf{R}}_s \nonumber \\
&={\bf{R}}_s^{1/2}[{\bf{R}}_s^{1/2}{\bf{H}}_1^{\rm{H}}{\bf{P}}^{\rm{H}}{\bf{H}}_2^{\rm{H}}({\bf{H}}_2{\bf{P}}{\bf{R}}_{n_1}
{\bf{P}}^{\rm{H}}{\bf{H}}_2^{\rm{H}}   +{\bf{R}}_{n_2})^{-1}\nonumber \\
&\times {\bf{H}}_2{\bf{P}}
{\bf{H}}_1{\bf{R}}_s^{1/2}+{\bf{I}}]^{-1}{\bf{R}}_s^{1/2}\triangleq {\boldsymbol \Psi}({\bf{P}}),
\end{align}based on which the determinant of the weighted MSE matrix is also reformulated as
\begin{align}
\label{Log_MSE}
{\rm{log}}|{\boldsymbol \Psi}({\bf{P}})|&={\rm{log}}|{\bf{R}}_s|-{\rm{log}}|{\bf{H}}_2{\bf{P}}
{\bf{H}}_1{\bf{R}}_s{\bf{H}}_1^{\rm{H}}{\bf{P}}^{\rm{H}}{\bf{H}}_2^{\rm{H}}\nonumber \\
&\times({\bf{H}}_2{\bf{P}}{\bf{R}}_{n_1}
{\bf{P}}^{\rm{H}}{\bf{H}}_2^{\rm{H}}   +{\bf{R}}_{n_2})^{-1}+{\bf{I}}|.
\end{align}
Therefore, based on (\ref{New_MSE_Matrix}) the optimization problem \textbf{P 1} given in (\ref{P_1}) can be rewritten as
\begin{align}
& {\min_{\bf{P}}} \ \ \ {\rm{Tr}}[{\boldsymbol \Psi}({\bf{P}})] \nonumber \\
& \ {\rm{s.t.}} \ \ \ {\rm{Tr}}({\bf{P}}({\bf{H}}_1{\bf{R}}_s{\bf{H}}_1^{\rm{H}}+{\bf{R}}_{n_1}){\bf{P}}^{\rm{H}}) \le P
\end{align} which is exactly the transceiver design of sum MSE minimization for dual-hop AF MIMO relaying systems, and has been separately discussed in  \cite{Schizas07} and \cite{Guan08}.

On the other hand, based on (\ref{Log_MSE}) the determinant minimization problem \textbf{P 2} given by (\ref{P_2}) is equivalent to
\begin{align}
& \max_{\bf{P}} \ \ {\rm{log}}|{\bf{R}}_s|-{\rm{log}}|{\boldsymbol \Psi}({\bf{P}})| \nonumber \\
& \ {\rm{s.t.}} \ \ \ {\rm{Tr}}({\bf{P}}({\bf{H}}_1{\bf{R}}_s{\bf{H}}_1^{\rm{H}}+{\bf{R}}_{n_1}){\bf{P}}^{\rm{H}}) \le P
\end{align}which is  the transceiver design of capacity maximization for dual-hop AF MIMO relaying systems, and has been separately discussed in \cite{Medina07} and \cite{Tang07}. As a result, based on these facts we have the following conclusion.

\noindent \underline{\textbf{Conclusion 4:}} In dual-hop AF MIMO relaying systems, the first hop can be understood as a weighting operation for the second hop as shown in Fig.~\ref{fig:1}. It may be the reason why the transceiver designs for AF MIMO relaying systems can enjoy almost all the results for point-to-point systems.

\section{Conclusions}
\label{conclusion}
In this letter, a matrix-field weighted MSE model was proposed, in which the weighting operation is extended to be a linear matrix function with respect to the original MSE matrix. Because for the proposed weighting operation all elements of the original MSE matrix are exploited, it can effectively reshape the data detection matrix and keep the all necessary information as well. Then the transceiver designs were formulated as an optimization problem aiming at minimizing an increasing matrix-monotone function of the weighted MSE matrix. The structure of the optimal solutions has been derived. Furthermore, two special cases of the optimization problems were discussed in more detail. It was also discovered that these two cases are exactly equivalent to the transceiver designs of sum MSE minimization and capacity maximization for dual-hop AF MIMO relaying systems.

\end{document}